# Calculation of double differential cross sections (DDCS) for H(3S) ionization using Bethe and Lewis Integrals

---

**Fahadul Islam*(ID), Sunil Dhar**

Department of Mathematics, Chittagong University of Engineering and Technology, Chittagong, Bangladesh.

Email: *fahadulislambgd@gmail.com, sdhar@cuet.ac.bd

**ABSTRACT**

Calculations have been made for the double differential cross section (DDCS) for the ionization of metastable hydrogen atoms in the 3S state by electron and positron impact at energies of 150 eV and 250 eV. The authors implemented the second Born approximation to the multiple scattering theory as their model, evaluated the corresponding analytical expressions using Bethe and Lewis integrals, and numerically computed them using MATLAB. The generated DDCS captures the features of both recoil and binary fragmentation and, at the same time, provides an overall qualitative consistency with earlier studies, although some differences can be found between them at certain emission angles. The present work, therefore, supplies new theoretical reference levels for ionization investigations in hydrogen-like systems, now that no experimental data are available for the metastable 3S state.

## 1. Introduction

An investigation of atomic ionization by charged projectiles, such as electrons or positrons, is one of the central fields in atomic physics. Calculating the precise values of these quantities for various types of cross sections–single, double [1], [2] and triple differential [3] for different choices of kinematic variables is an attractive and fascinating subject for applied mathematics. In the past fifty years, new experimental results have been added to this line of research, including astrophysics, plasma physics and Fusion technologies. This kind of research provides information on the ionization of metastable state [4] - [19] hydrogen atoms. Moreover, it is expected that experimental information on this area will be published. Our present calculations obey the Lewis integral [20].

The Double Differential Cross Section (DDCS) indicates the spread in energy and angle in which the secondary electrons are produced in atomic ionization collisions. The DDCS data are helpful for the study of astrophysical and upper atmospheric events, electron impact spectra, and the study of secondary effects produced by slow secondary electrons and many other phenomena discussed by Das et al. [1]. Shyn [2] for the helium atom, which is the least troublesome atom from an experimental point of view, essentially just computed available DDCS experimental data in angle and energy. The ionization of fast particles, Bethe [3], was the first to investigate, non-relativistic ally, the process of ionization by fast particles using quantum mechanics.

A multi-scattering theory [3] has been used in calculating the DDCS of ionization of metastable 3S-state hydrogen atoms by 150 eV and 250 eV electron impact. The multi-scattering wave function [3], and [4] has been expressed for two electrons in a Coulomb field, taking into account higher-order effects and correlation effects. Using these wave functions, the TDCS has been computed with great success for various kinematical conditions for electron-hydrogen ionization collisions, both for the ground state and for the metastable 3S-state at non-relativistic energies [3], and [4]. The concept proposed by Das et al. [3] is very modest and easily applicable to the present study of DDCS of ionization of the metastable 3S-state of the hydrogen atom by electrons at intermediate energies.

The DDCS results were obtained as integrals of TDCS results [9], [10], [11], [12], and [13] over scattered electron directions and compared with the measurements of Shyn [2] and the predictions of Das et al. [1]. One more integration to extract the DDCS results and compare them with the measurements of Shyn [2] and Das et al. [1]. It will be interesting and relevant to also use the wave function of Das et al. [3], and [4] in the present work for the ionization of metastable 3S-state hydrogen atoms by electrons.

## 2. Theoretical Method

Ionization cross-sections are based on the number of ionizations per unit time and per unit target to the incident electron flux. The most detailed knowledge exists to date about the single ionization processes of the following type:

$$e^- + H(3S) \rightarrow H^+ + 2e^- \qquad \textit{for electron impact} \tag{1}$$

$$p^- + H(3S) \rightarrow H^+ + 2e^- \qquad \textit{for positron impact} \tag{2}$$

Where the notations are given in the caption and have been extracted in the coplanar geometry by discussing the triple differential cross sections (TDCS) measured in (e, 2e) coincidence experiments. For the ionization of a hydrogen atom by electrons [3], the direct T-matrix element [4] can be expressed by,

$$T_{FI}=\langle\Psi_F^{(-)}\left(\bar{\gamma}_a,\bar{\gamma}_b\right)|V_I\left(\bar{\gamma}_a,\bar{\gamma}_b\right)|\Phi_I\left(\bar{\gamma}_a,\bar{\gamma}_b\right)\rangle \tag{3}$$

Here the perturbation potential $\mathrm{V_I}\left(\bar{\gamma}_\mathrm{a},\bar{\gamma}_\mathrm{b}\right)$ is given by

$$V_I\left(\bar{\gamma}_a,\bar{\gamma}_b\right)=\frac{Z}{\gamma_b}-\frac{Z}{\gamma_{ab}} \tag{4}$$

The nuclear charge of the hydrogen atom is Z= -1 for electrons and Z= +1 for positrons, $\bar{\gamma}_\mathrm{a}$ and $\bar{\gamma}_\mathrm{b}$ are the distances of the two electrons from the nucleus and $\gamma_{ab}$ is the distance between the two electrons. The initial channel unperturbed wave function is

$$\Phi_I\left(\bar{\gamma}_a,\bar{\gamma}_b\right)=\frac{e^{i\cdot\bar{p}_i\cdot\bar{\gamma}_b}}{(2\pi)^{\frac{3}{2}}}\phi_{3S}\left(\bar{\gamma}_a\right)=\frac{e^{i\cdot\bar{p}_i\cdot\bar{\gamma}_b}}{(2\pi)^{\frac{3}{2}}}\cdot\frac{1}{81\sqrt{3\pi}}\left(27\text{-}18\gamma_a+2\gamma_a^2\right)e^{-\lambda_a\gamma_a} \tag{5}$$

Here

$$\phi_{3S}\left(\bar{\gamma}_a\right)=\frac{1}{81\sqrt{3\pi}}\left(27\text{-}18\gamma_a+2\gamma_a^2\right)e^{-\lambda_a\gamma_a} \tag{6}$$

Here $\lambda_a=\frac{1}{3}$ , $\phi_{3S}(\bar{\gamma}_a)$ is the hydrogen 3S-state wave function, and $\Psi_F^{(-)}\left(\bar{\gamma}_a,\bar{\gamma}_b\right)$is the final three-particle scattering state wave function [3] with the electrons being in the continuum with momenta $\bar{p}_a$ *and* $\bar{p}_b$. And the coordinates of the two electrons are $\bar{\gamma}_a$ *and* $\bar{\gamma}_b$ respectively. Here the approximate wave function $\Psi_F^{(-)}\left(\bar{\gamma}_a,\bar{\gamma}_b\right)$ [4] is given by

$$\Psi_F^{(-)}\left(\bar{\gamma}_a,\bar{\gamma}_b\right)=\frac{N\left(\bar{p}_a,\bar{p}_b\right)\left[\phi_{\bar{p}_a}^{(-)}\left(\bar{\gamma}_a\right)e^{i\bar{p}_b\cdot\bar{\gamma}_b}+\phi_{\bar{p}_2}^{(-)}\left(\bar{\gamma}_b\right)e^{i\bar{p}_a\cdot\bar{\gamma}_a}+\phi_{\bar{p}}^{(-)}\left(\bar{\gamma}\right)e^{i\bar{P}.\bar{R}}\text{-}2e^{i\bar{p}_a\cdot\bar{\gamma}_a+i\bar{p}_b\cdot\bar{\gamma}_b}\right]}{(2\pi)^3} \tag{7}$$

*Here*

$$\bar{\gamma}=\frac{\bar{\gamma}_b-\bar{\gamma}_a}{2},\quad \bar{R}=\frac{\bar{\gamma}_b+\bar{\gamma}_a}{2},\quad \bar{p}=\left(\bar{p}_b\text{-}\bar{p}_a\right),\ \bar{P}=\left(\bar{p}_b+\bar{p}_a\right)$$

The scattering amplitude [4] may be written as

$$F\left(\bar{p}_a,\bar{p}_b\right)=N\left(\bar{p}_a,\bar{p}_b\right)[F_{eT}+F_{PT}+F_{Pe}-2F_{PWB}] \tag{8}$$

Where $F_{eT}, F_{PT}, F_{Pe}$ *and* $F_{PWB}$ are the amplitudes corresponding to the four terms of Eq. (7) respectfully. Here $N(\bar{p}_a,\bar{p}_b)$ is the normalization constant, given by,

$$\left|N(\bar{p}_a,\bar{p}_b)\right|^{-2}=\left|7\text{-}2[\lambda_a+\lambda_b+\lambda_c]\text{-}\left[\frac{2}{\lambda_a}+\frac{2}{\lambda_b}+\frac{2}{\lambda_c}\right]+\left[\frac{\lambda_a}{\lambda_b}+\frac{\lambda_a}{\lambda_c}+\frac{\lambda_b}{\lambda_a}+\frac{\lambda_b}{\lambda_c}+\frac{\lambda_c}{\lambda_a}+\frac{\lambda_c}{\lambda_b}\right]\right| \tag{9}$$

Where

$\lambda_a=e^{\frac{\pi\alpha_a}{2}}\Gamma(1\text{-}i\alpha_a),\ \ \alpha_a=\frac{1}{p_a};$

$\lambda_b=e^{\frac{\pi\alpha_b}{2}}\Gamma(1\text{-}i\alpha_b),\ \ \alpha_b=\frac{1}{p_b};$

$\lambda_c=e^{\frac{\pi\alpha}{2}}\Gamma(1\text{-}i\alpha),\ \ \alpha=\text{-}\frac{1}{p}$ .

Here $\phi_{\bar{q}}^{(-)}(\bar{\gamma})$ is the coulombs wave function, given by,

$$\phi_{\bar{q}}^{(-)}(\bar{\gamma})=e^{\frac{\pi\alpha}{2}}\Gamma(1+i\alpha)e^{i\bar{q}\cdot\bar{\gamma}}\ \ {}_1F_1(\text{-}i\alpha,1,\text{-}i[q\gamma+\bar{q}.\bar{\gamma}]) \tag{10}$$

The general one-dimensional integral representation of the confluent hyper geometric function is written by,

$${}_1F_1(a,c,z)\ =\ \frac{\Gamma(c)}{(a)\Gamma(c-a)}\int_0^1 dx\ x^{(a-1)}(1-x)^{(c-a-1)}e^{(xz)} \tag{11}$$

For the electron impact ionization, the parameters $\alpha_a, \alpha_b$ *and* $\alpha$ are given by,

$$\alpha_a\ =\frac{1}{P_a}\ for\ \bar{q}\ =\bar{p}_a\ ,\ \alpha_b\ =\frac{1}{p_b}\ for\ \ \bar{q}\ =\bar{p}_b\ \ and\ \ \ \alpha\ =-\frac{1}{p}\ for\ \ \bar{q}\ =\bar{p}.$$

Now applying equations (5) and (7) to the equation (3), we get

$$\mathrm{T_{FI}}\ =\mathrm{N}(\bar{\mathrm{p}}_\mathrm{a},\bar{\mathrm{p}}_\mathrm{b})[\mathrm{T_B}+\mathrm{T_{B'}}+\mathrm{T_I}-2\mathrm{T_{PB}}] \tag{12}$$

Where

$$T_B\ =\langle\Phi_{\bar{p}_a}^{(-)}(\bar{\gamma}_a)\ e^{i\bar{p}_b\cdot\bar{\gamma}_b}|\ V_I\ |\ \Phi_I(\bar{\gamma}_a,\bar{\gamma}_b)\rangle \tag{13}$$

$$T_{B'}\ =\langle\Phi_{\bar{p}_b}^{(-)}(\bar{\gamma}_b)\ e^{i\bar{p}_a\cdot\bar{\gamma}_a}|\ V_I\ |\ \Phi_I(\bar{\gamma}_a,\bar{\gamma}_b)\rangle \tag{14}$$

$$T_I = \langle \Phi_{\bar{p}}^{(-)}(\bar{\gamma})\, e^{i.\bar{P}.\bar{R}} \mid V_I \mid \Phi_I(\bar{\gamma}_a, \bar{\gamma}_b) \rangle \tag{15}$$

$$T_{PB} = \langle e^{i\bar{p}_a.\bar{\gamma}_a + i\bar{p}_b.\bar{\gamma}_b} \mid V_I \mid \Phi_I(\bar{\gamma}_a, \bar{\gamma}_b) \rangle \tag{16}$$

For the first-born approximation, equation (13) may be written as

$$T_B = \frac{1}{162\sqrt{6}\pi^2} \langle \Phi_{\bar{p}_a}^{(-)}(\bar{\gamma}_a)\, e^{i.\bar{p}_b.\bar{\gamma}_b} \left| \frac{1}{\gamma_{ab}} - \frac{1}{\gamma_b} \right| e^{i.\bar{p}_i.\bar{\gamma}_b} \,(27 - 18\gamma_a + 2\gamma_a^2)\, e^{-\lambda_a.\gamma_a} \rangle$$

$$T_B = T_{B_1} + T_{B_2} + T_{B_3} + T_{B_4} + T_{B_5} + T_{B_6} \tag{17}$$

Where

$$T_{B_1} = \frac{1}{6\sqrt{6}\pi^2} \int \phi_{\bar{p}_a}^{(-)*}(\bar{\gamma}_a)\, e^{-i.\bar{p}_b.\bar{\gamma}_b} \frac{1}{\gamma_{ab}}\, e^{i\bar{p}_i.\bar{\gamma}_b}\, e^{-\lambda_a \gamma_a}\, d^3\gamma_a\, d^3\gamma_b \tag{18}$$

$$T_{B_2} = -\frac{1}{9\sqrt{6}\pi^2} \int \phi_{\bar{p}_a}^{(-)*}(\bar{\gamma}_a)\, e^{-i.\bar{p}_b.\bar{\gamma}_b} \frac{\gamma_a}{\gamma_{ab}}\, e^{i.\bar{p}_i.\bar{\gamma}_b}\, e^{-\lambda_a \gamma_a}\, d^3\gamma_a\, d^3\gamma_b \tag{19}$$

$$T_{B_3} = \frac{1}{81\sqrt{6}\pi^2} \int \phi_{\bar{p}_a}^{(-)*}(\bar{\gamma}_a)\, e^{-i.\bar{p}_b.\bar{\gamma}_b} \frac{{\gamma_a}^2}{\gamma_{ab}}\, e^{i.\bar{p}_i.\bar{\gamma}_b}\, e^{-\lambda_a \gamma_a}\, d^3\gamma_a\, d^3\gamma_b \tag{20}$$

$$T_{B_4} = -\frac{1}{6\sqrt{6}\pi^2} \int \phi_{\bar{p}_a}^{(-)*}(\bar{\gamma}_a)\, e^{-i.\bar{p}_b.\bar{\gamma}_b} \frac{1}{\gamma_b}\, e^{i.\bar{p}_i.\bar{\gamma}_b}\, e^{-\lambda_a \gamma_a}\, d^3\gamma_a\, d^3\gamma_b \tag{21}$$

$$T_{B_5} = \frac{1}{9\sqrt{6}\pi^2} \int \phi_{\bar{p}_a}^{(-)*}(\bar{\gamma}_a)\, e^{-i.\bar{p}_b.\bar{\gamma}_b} \frac{\gamma_a}{\gamma_b}\, e^{i.\bar{p}_i.\bar{\gamma}_b}\, e^{-\lambda_a \gamma_a}\, d^3\gamma_a\, d^3\gamma_b \tag{22}$$

$$T_{B_6} = -\frac{1}{81\sqrt{6}\pi^2} \int \phi_{\bar{p}_a}^{(-)*}(\bar{\gamma}_a)\, e^{-i.\bar{p}_b.\bar{\gamma}_b} \frac{{\gamma_a}^2}{\gamma_b}\, e^{i.\bar{p}_i.\bar{\gamma}_b}\, e^{-\lambda_a \gamma_a}\, d^3\gamma_a\, d^3\gamma_b \tag{23}$$

Here $T_{B_4} = 0$ *and* $T_{B_5} = 0$, (for orthogonality condition)

Since electron-nucleus interaction $\frac{1}{\gamma_b}$ does not contribute to first-born term; because of the orthogonality of the initial and final target states.

The above equations may be written by using Bethe integral [3], as

$$T_{B_1} = \frac{4e^{\left(\frac{\pi\alpha_a}{2}\right)} \Gamma(1 - i\alpha_a)\left(\bar{t}^2 - \bar{p}_a.\bar{t} - i\alpha_a\, \bar{p}_a.\bar{t}\right) e^{(i\alpha_a \ln \omega)}}{3\sqrt{6} t^2 \{\bar{t}^2 - (i\lambda_a + \bar{p}_a)^2\} \{\lambda_a^2 + (\bar{t} - \bar{p}_a)^2\}^{2 - i\alpha_a}} \tag{24}$$

$$T_{B_2} = \frac{16\sqrt{2}e^{\left(\frac{\pi\alpha_a}{2}\right)}\Gamma(1-i\alpha_a)\left(\bar{t}^2-\bar{p}_a.\bar{t}-i\alpha_a\,\bar{p}_a.\bar{t}\right)e^{(i\alpha_a\ln\omega)}}{3^{\frac{5}{2}}t^2\{\bar{t}^2-(i\lambda_a+\bar{p}_a)^2\}\left\{\lambda_a^2+(\bar{t}-\bar{p}_a)^2\right\}^2} \tag{25}$$

$$= \left[\frac{1}{\lambda_a} - \frac{4\lambda_a}{{\lambda_a}^2+(\bar{p}_a-\bar{t})^2} - \frac{2\,\alpha_a\,\bar{p}_a\left(\lambda_a^2-p_a^2+t^2\right)}{\left(\lambda_a^2-p_a^2+t^2\right)^2+4\lambda_a^2\,p_a^2} + \frac{4\alpha_a\lambda_a^2\,p_a}{\left(\lambda_a^2-p_a^2+t^2\right)^2+4\lambda_a^2\,p_a^2}\right]$$

$$+i\left[\frac{2\lambda_a\,\alpha_a}{\lambda_a^2+(\bar{p}_a-\bar{t})^2} - \frac{4\alpha_a\lambda_a p_a^2}{\left(\lambda_a^2-p_a^2+t^2\right)^2+4\lambda_a^2\,p_a^2} - \frac{2\,\lambda_a\,\alpha_a\left(\lambda_a^2-p_a^2+t^2\right)}{\left(\lambda_a^2-p_a^2+t^2\right)^2+4\lambda_a^2\,p_a^2}\right]$$

$$T_{B_3} = \frac{4e^{\left(\frac{\pi\alpha_a}{2}\right)}\Gamma(1-i\alpha_a)\left(\bar{t}^2-\bar{p}_a.\bar{t}-i\alpha_a\,\bar{p}_a.\bar{t}\right)e^{(i\alpha_a\ln\omega)}}{81\sqrt{2}\pi t^2\{\bar{t}^2-(i\lambda_a+\bar{p}_a)^2\}\left\{\lambda_a^2+(\bar{t}-\bar{p}_a)^2\right\}^{1+\alpha_a}} \tag{26}$$

$$T_{B_4} = 0 \tag{27}$$

$$T_{B_5} = 0 \tag{28}$$

$$T_{B_6} = -\frac{8e^{\left(\frac{\pi\alpha_a}{2}\right)}\Gamma(1-i\alpha_a)\left(\bar{t}^2-\bar{p}_a.\bar{t}-i\alpha_a\,\bar{p}_a.\bar{t}\right)e^{(i\alpha_a\ln\omega)}}{81\sqrt{2}\pi t^2\{\bar{t}^2-(i\lambda_a+\bar{p}_a)^2\}\left\{\lambda_a^2+(\bar{t}-\bar{p}_a)^2\right\}^{2-i\alpha_a}} \tag{29}$$

*Here*

$$\omega = \frac{\lambda_a^2+(\bar{t}-\bar{p}_a)^2}{\bar{t}^2-(i\lambda_a+\bar{p}_a)^2} \;\; \textit{with}\; \bar{t} = \bar{p}_i - \bar{p}_b.\; \bar{p}_a.\bar{t} = \bar{p}_a.(\bar{p}_i - \bar{p}_b)$$

$$= \bar{p}_a.\bar{p}_i - \bar{p}_a.\bar{p}_b \;= p_a p_i \cos\theta_a - p_a p_b \cos\theta_{ab}$$

$Here\;\; \cos\theta_{ab} = \cos\theta_a \cos\theta_b - \sin\theta_a\; \sin\theta_b$ For phase ξ value we have

$$\Gamma(1+i\alpha_a) = e^{i\xi}|\Gamma(1+i\alpha_a)|,\;\; \Gamma(1-i\alpha_a) = e^{-i\xi}|\Gamma(1-i\alpha_a)|$$

$$|\Gamma(1-i\alpha_a)| = \sqrt{\Gamma(1-i\alpha_a)\,\Gamma(1+i\alpha_a)} = \sqrt{\frac{i\alpha_a\pi}{\sin(i\alpha_a\pi)}} \;=\; \sqrt{\frac{i\alpha_a\pi}{\mathrm{isinh}(i\alpha_a\pi)}}$$

$$|\Gamma(1-i\alpha_a)| = \sqrt{\frac{2\,\pi\,\alpha_a}{e^{\pi\,\alpha_a}-e^{-\pi\,\alpha_a}}}$$

The second term of equation (14) can be written as follows:

$$\mathrm{T}_{\mathrm{B}'} = \frac{1}{162\sqrt{6}\pi^2}\langle\Phi_{\bar{\mathrm{p}}_\mathrm{b}}^{(-)}(\bar{\gamma}_\mathrm{b})e^{\mathrm{i}\bar{\mathrm{p}}_\mathrm{a}.\bar{\gamma}_\mathrm{a}}\left|\frac{1}{\gamma_{ab}} - \frac{1}{\gamma_b}\right|\Phi_\mathrm{i}(\bar{\gamma}_\mathrm{a},\bar{\gamma}_\mathrm{b})\rangle$$

$$= \frac{1}{162\sqrt{6}\pi^2}\int \Phi_{\bar{p}_b}^{(-)*}(\bar{\gamma}_b)e^{-i\bar{p}_a\cdot\bar{\gamma}_a}\left|\frac{1}{\gamma_{ab}}-\frac{1}{\gamma_b}\right| e^{i\bar{p}_i\cdot\bar{\gamma}_b}(27-18\gamma_a+2\gamma_a^2)e^{-\lambda_a\gamma_a}d^3\gamma_a d^3\gamma_b$$

$$T_{B'} = T_{B_1}{}' + T_{B_2}{}' + T_{B_3}{}' + T_{B_4}{}' + T_{B_5}{}' + T_{B_6}{}' \tag{30}$$

Where

$$T_{B_1}{}' = \frac{1}{6\sqrt{6}\pi^2}\int \Phi_{\bar{p}_b}^{(-)*}(\bar{\gamma}_b)e^{-i\bar{p}_a\cdot\bar{\gamma}_a}\frac{1}{\gamma_{ab}}e^{i\bar{p}_i\cdot\bar{\gamma}_b}e^{-\lambda_a\gamma_a}d^3\gamma_a d^3\gamma_b \tag{31}$$

$$T_{B_2}{}' = -\frac{1}{9\sqrt{6}\pi^2}\int \Phi_{\bar{p}_b}^{(-)*}(\bar{\gamma}_b)e^{-i\bar{p}_a\cdot\bar{\gamma}_a}\frac{\gamma_a}{\gamma_{ab}}\ e^{i\bar{p}_i\cdot\bar{\gamma}_b}e^{-\lambda_a\gamma_a}d^3\gamma_a d^3\gamma_b \tag{32}$$

$$T_{B_3}{}' = \frac{1}{81\sqrt{6}\pi^2}\int \Phi_{\bar{p}_b}^{(-)*}(\bar{\gamma}_b)e^{-i\bar{p}_a\cdot\bar{\gamma}_a}\frac{\gamma_a^2}{\gamma_{ab}}\ e^{i\bar{p}_i\cdot\bar{\gamma}_b}e^{-\lambda_a\gamma_a}d^3\gamma_a d^3\gamma_b \tag{33}$$

$$T_{B_4}{}' = -\frac{1}{6\sqrt{6}\pi^2}\int \Phi_{\bar{p}_b}^{(-)*}(\bar{\gamma}_b)e^{-i\bar{p}_a\cdot\bar{\gamma}_a}\frac{1}{\gamma_b}e^{i\bar{p}_i\cdot\bar{\gamma}_b}e^{-\lambda_a\gamma_a}d^3\gamma_a d^3\gamma_b \tag{34}$$

$$T_{B_5}{}' = \frac{1}{9\sqrt{6}\pi^2}\int \Phi_{\bar{p}_b}^{(-)*}(\bar{\gamma}_b)e^{-i\bar{p}_a\cdot\bar{\gamma}_a}\frac{\gamma_a}{\gamma_b}\ e^{i\bar{p}_i\cdot\bar{\gamma}_b}\ e^{-\lambda_a\cdot\gamma_a}d^3\gamma_a d^3\gamma_b \tag{35}$$

$$T_{B_6}{}' = -\frac{1}{81\sqrt{6}\pi^2}\ \int \Phi_{\bar{p}_b}^{(-)*}(\bar{\gamma}_b)e^{-i\bar{p}_a\cdot\bar{\gamma}_a}\frac{{\gamma_a}^2}{\gamma_b}e^{i\bar{p}_i\cdot\bar{\gamma}_b}\ e^{-\lambda_a\cdot\gamma_a}d^3\gamma_a d^3\gamma_b \tag{36}$$

Using Coulomb wave function given earlier the above equation (31) is reduced to

$$T_{B_1}{}' = \frac{8ie^{\frac{\pi\alpha_b}{2}}\Gamma(1-i\alpha_b)}{6\sqrt{6}\pi^2\Gamma(i\alpha_b)\Gamma(1-i\alpha_b)}\int_0^1 dx\, x^{i\alpha_b-1}(1-x)^{-i\alpha_b}\,\phi(x) \tag{37}$$

Where

$$\phi(x) = \frac{\partial^2}{\partial\lambda_a\partial\lambda_b}\ I(\lambda_c,\lambda_a,\bar{q}_a',\bar{q}_b'),\ \ \phi(x) = \frac{\partial^2}{\partial\lambda_a\partial\lambda_b}\left(\frac{2\pi^2}{\lambda_a\beta}\right)$$

Where $\beta = {\bar{q}_b'}^2 - k^2 - ik\mu_a z_a$

From Lewis integral [20] we have,

$$I(\lambda_c,\lambda_a,\bar{q}_a',\bar{q}_b') = \int \frac{d^3q}{(q^2-k^2-\lambda\varepsilon)\left[(\bar{q}-\bar{q}_a)^2+{\lambda_a}^2\right](\bar{q}-\bar{q}_b)^2} = \frac{8\pi\lambda_a}{\left\{{\lambda_b}^2+(\bar{q}-p_i+p_b-xp_b)^2\right\}^2} \tag{38}$$

Here $I(\lambda_c,\lambda_a,\bar{q}_a',\bar{q}_b')$ is the Lewis Integral [20] and is given by

$$I(\lambda_c, \lambda_a, \bar{q}'_a, \bar{q}'_b) = \int \frac{d^3\bar{q}'}{(\bar{q}'-\bar{q}_b{}')^2(\bar{q}'^2-\lambda_c{}^2-i\varepsilon)\left[\lambda_a{}^2+(\bar{q}'-\bar{q}_b{}')^2\right]} \qquad (39)$$

$\bar{q}'_b = \bar{p}_b - \bar{p}_a - x\bar{p}_b$, $\bar{q}'_a = \bar{p}_a + \bar{p}_b - \bar{p}_l - x\bar{p}_b$ , $\lambda_a = \frac{1}{3}$

$\lambda_c = x\bar{p}_b$ and $\varepsilon$ is constant. By using Bethe Integral [3] $T_{B_1{}'}$ of equation (37) becomes,

$$T_{B_1{}'} = -\frac{8ie^{\frac{\pi\alpha_b}{2}}\Gamma(1-i\alpha_b)}{27\sqrt{6}\pi^2\Gamma(i\alpha_b)\Gamma(1-i\alpha_b)}\left[\frac{\sinh\pi\alpha_b}{\pi}\int_0^1 dx\left(\frac{x}{1-x}\right)^{i\alpha_b}\left\{\frac{\phi(x)-\phi(o)}{x}\right\} - i\phi(o)\right] \qquad (40)$$

where $\left(\frac{x}{1-x}\right)^{i\alpha_b} = e^{i\alpha_b \log\left(\frac{x}{1-x}\right)} = \cos\left\{\alpha_b \log\left(\frac{x}{1-x}\right)\right\} + i\sin\left\{\alpha_b \log\left(\frac{x}{1-x}\right)\right\}$

Differentiating equation (40) with respect to $\lambda_a$, we get the term$T_{B_1{}'}$. Similarly, using the

Bethe Integral [3] the term $T_{B_2{}'}$ of the equation (32) becomes

$$T_{B_2{}'} = -\frac{32\lambda_a\Gamma(1-i\alpha_b)}{9\sqrt{6}\left(\lambda_a{}^2+p_a{}^2\right)^{2-i\alpha_b}(t^2)^{1+i\alpha_b}} \qquad (41)$$

Where

$$\frac{\lambda_a}{\left(\lambda_a{}^2+p_a{}^2\right)^{2-i\alpha_b}} = \frac{\lambda_a}{\left(\lambda_a{}^2+p_a{}^2\right)^2} * \left(\cos\left(\alpha_b \ln\left(\lambda_a{}^2+p_a{}^2\right)\right) + i\sin\left(\alpha_b \ln\left(\lambda_a{}^2+p_a{}^2\right)\right)\right)$$

Differentiating (41) with respect to $\lambda_a$ we find the term $T_{B_2{}'}$ .

Similarly using Bethe Integral [3] for evaluating the term of equation (33) is $T_{B_3{}'}$ as

$$T_{B_3{}'} = -\frac{8\sqrt{2}\,ie^{\frac{\pi\alpha_b}{2}}\Gamma(1-i\alpha_b)}{3\sqrt{3}\,\Gamma(i\alpha_b)\Gamma(1-i\alpha_b)}\left[\int_0^1 dx\left(\frac{x}{1-x}\right)^{i\alpha_b}\phi_a(x)\right] \qquad (42)$$

We used here, $\Gamma(z)\Gamma(1-z) = \frac{\pi}{\sin\pi z}$

After differentiating the equation (42) two times with a respect to $\lambda_a$ we have the term $T_{B_3{}'}$.

Bethe Integral [3] makes the term of equation (34) is $T_{B_4{}'}$ as

$$T_{B_4'} = \frac{8\sqrt{2}\, e^{\frac{\pi\alpha_b}{2}} \lambda_a \Gamma(1-i\alpha_b)}{3\sqrt{3}\left(\lambda_a{}^2+p_a{}^2\right)^{2-i\alpha_b}(t^2)^{1+i\alpha_b}} = \frac{4\lambda_a}{\lambda_a{}^2+p_a{}^2} - i\,\frac{2\lambda_a\alpha_b}{\lambda_a{}^2+p_a{}^2} \tag{43}$$

After differentiating the above equation (43) with a respect to $\lambda_a$ we get the term $T_{b_4'}$. Similarly, Bethe Integral [3] makes the term of equation (35) $T_{B_5'}$ as

$$T_{B_5'} = -\frac{32\, ie^{\frac{\pi\alpha_b}{2}}\Gamma(1-i\alpha_b)}{81\sqrt{6}\pi^2\Gamma(i\alpha_b)\Gamma(1-i\alpha_b)}\left[\int_0^1 dx\left(\frac{x}{1-x}\right)^{i\alpha_b} \phi_b(x)\right] \tag{44}$$

$$= \frac{32\, ie^{\frac{\pi\alpha_b}{2}}}{81\sqrt{6}\pi^3}\left[\frac{e^{\frac{\pi\alpha_b}{2}} - e^{\frac{-\pi\alpha_b}{2}}}{2}\right]\phi_b(x) + i\,\frac{32\, ie^{\frac{\pi\alpha_b}{2}}}{81\sqrt{6}\pi^3}\;\phi_b(0)$$

After differentiating the equation (44) with a respect to $\lambda_a$ we have the term $T_{B_5'}$. Again, doing the calculation using Bethe Integral [3] makes the term of equation (36) is $T_{B_6'}$ becomes

$$T_{B_6'} = -\frac{32\,\lambda_a\Gamma(1-i\alpha_b)}{81\sqrt{6}\left(\lambda_a{}^2+p_a{}^2\right)^{2-i\alpha_b}(t^2)^{1+i\alpha_b}} \tag{45}$$

$$= \frac{1}{\lambda_a} - \frac{4\lambda_a}{\lambda_a{}^2+p_a{}^2} + i\,\frac{2\lambda_a\alpha_b}{\lambda_a{}^2+p_a{}^2}$$

After differentiating the equation (44) two times with a respect to $\lambda_a$ we have the term $T_{B_6'}$. Then putting the values of $T_{B_1'}$, $T_{B_2'}$, $T_{B_3'}$, $T_{B_4'}$, $T_{B_5'}$ and $T_{B_6'}$ in the equation (30) we get $T_{B'}$. The third term of equation (15) can be written as follows:

$$T_I = \frac{1}{162\sqrt{6}\pi^2}\langle \Phi_{\bar{p}}^{(-)*}(\bar{\gamma})e^{i\bar{P}.\bar{R}}\left|\frac{1}{\gamma_{ab}} - \frac{1}{\gamma_b}\right|(27-18\gamma_a+2\gamma_a^2)e^{i\bar{p}_i\cdot\bar{\gamma}_b}e^{-\lambda_a\gamma_a}\rangle$$

$$T_I = \frac{1}{162\sqrt{6}\pi^2}\int \Phi_{\bar{p}}^{(-)*}(\bar{\gamma})e^{i\bar{P}.\bar{R}}\left|\frac{1}{\gamma_{ab}} - \frac{1}{\gamma_b}\right|(27-18\gamma_a+2\gamma_a^2)e^{i\bar{p}_i\cdot\bar{\gamma}_b}e^{-\lambda_a\gamma_a}\, d^3\gamma_a d^3\gamma_b$$

$$T_I = T_{I_1} + T_{I_2} + T_{I_3} + T_{I_4} + T_{I_5} + T_{I_6} \tag{46}$$

where

$$T_{I_1} = \frac{1}{6\sqrt{6}\pi^2} \int \Phi_{\bar{p}}^{(-)*}(\bar{\gamma}) e^{i\bar{P}.\bar{R}} \frac{1}{\gamma_{ab}} e^{i\bar{p}_i \cdot \bar{\gamma}_b} e^{-\lambda_a \gamma_a} d^3\gamma_a d^3\gamma_b \tag{47}$$

$$T_{I_2} = -\frac{1}{9\sqrt{6}\pi^2} \int \Phi_{\bar{p}}^{(-)*}(\bar{\gamma}) e^{i\bar{P}.\bar{R}} \frac{\gamma_a}{\gamma_{ab}} e^{i\bar{p}_i \cdot \bar{\gamma}_b} e^{-\lambda_a \gamma_a} d^3\gamma_a d^3\gamma_b \tag{48}$$

$$T_{I_3} = \frac{1}{81\sqrt{6}\pi^2} \int \Phi_{\bar{p}}^{(-)*}(\bar{\gamma}) e^{i\bar{P}.\bar{R}} \frac{\gamma_a^2}{\gamma_{ab}} e^{i\bar{p}_i \cdot \bar{\gamma}_b} e^{-\lambda_a \gamma_a} d^3\gamma_a d^3\gamma_b \tag{49}$$

$$T_{I_4} = -\frac{1}{6\sqrt{6}\pi^2} \int \Phi_{\bar{p}}^{(-)*}(\bar{\gamma}) e^{i\bar{P}.\bar{R}} \frac{1}{\gamma_b} e^{i\bar{p}_i \cdot \bar{\gamma}_b} e^{-\lambda_a \gamma_a} d^3\gamma_a d^3\gamma_b \tag{50}$$

$$T_{I_5} = \frac{1}{9\sqrt{6}\pi^2} \int \Phi_{\bar{p}}^{(-)*}(\bar{\gamma}) e^{i\bar{P}.\bar{R}} \frac{\gamma_a}{\gamma_b} e^{i\bar{p}_i \cdot \bar{\gamma}_b} e^{-\lambda_a \cdot \gamma_a} d^3\gamma_a d^3\gamma_b \tag{51}$$

$$T_{I_6} = -\frac{1}{81\sqrt{6}\pi^2} \int \Phi_{\bar{p}}^{(-)*}(\bar{\gamma}) e^{i\bar{P}.\bar{R}} \frac{{\gamma_a}^2}{\gamma_b} e^{i\bar{p}_i \cdot \bar{\gamma}_b} e^{-\lambda_a \cdot \gamma_a} d^3\gamma_a d^3\gamma_b \tag{52}$$

Using the Bethe Integral [3] we get from equation (47)

$$T_{I_1} = \frac{8\pi\lambda_a i e^{\frac{\pi\alpha_b}{2}} \Gamma(1-i\alpha_b) 4\pi}{6\sqrt{6}\left\{{\lambda_a}^2+(\bar{t}-\bar{p}_a)^2\right\}^2 (t^2)^{1+i\alpha} \left(p_i^2+\bar{p}_i.\bar{p}_a+\bar{p}_i.\bar{p}_b+\bar{p}_a.\bar{p}_b\right)^{-i\alpha}} \tag{53}$$

$$= \frac{16\, e^{\frac{\pi\alpha}{2}} \Gamma(1-i\alpha) e^{(i\alpha \ln\theta)}}{3\sqrt{6}t^2} * \frac{\lambda_1}{\left\{\lambda_a^2+(\bar{t}-\bar{p})^2\right\}^2}$$

Differentiating the above equation (53) with a respect to $\lambda_a$ we have $T_{I_1}$

Again, using Bethe Integral [3] equation (48) can be written as

$$T_{I_2} = \pm \frac{4\sqrt{2} i e^{\frac{\pi\alpha_b}{2}}}{3\sqrt{3}\pi^2 \Gamma(i\alpha)\Gamma(1-i\alpha)} \left[\int_0^1 dx\, x^{i\alpha}(1-x)^{-i\alpha} \left\{\frac{\phi(x)-\phi(o)}{x}\right\} - i\phi(o)\right] \tag{54}$$

Differentiating the above equation (54) with a respect to $\lambda_a$ we have the $T_{I_2}$

Using Bethe Integral [3] equation (49) can be written as,

$$T_{I_3} = \frac{8\pi\lambda_a i e^{\frac{\pi\alpha_2}{2}}\Gamma(1-i\alpha_2)4\pi}{9\sqrt{6}\{\lambda_a{}^2+(\bar{t}-\bar{p}_a)^2\}^2(t^2)^{1+i\alpha}\left(p_i^2+\bar{p}_i.\bar{p}_a+\bar{p}_i.\bar{p}_b+\bar{p}_a.\bar{p}_b\right)^{-i\alpha}} \tag{55}$$

Differentiating the above equation (55) with a respect to $\lambda_a$ we have the $T_{i_3}$

Again, using Bethe Integral [3] equation (50) can be written as

$$T_{I_4} = \pm\frac{ie^{\frac{\pi\alpha_b}{2}}(e^{\pi\alpha}-e^{-\pi\alpha})}{9\sqrt{6}\pi^2\Gamma(i\alpha)\Gamma(1-i\alpha)}\left[\int_0^1 dx\, x^{i\alpha}(1-x)^{-i\alpha}\Gamma(1-i\alpha)\left\{\frac{\phi_a(x)-\phi_a(0)}{x}\right\} - i\Gamma(1-i\alpha)\phi_a(0)\right] \tag{56}$$

Differentiating the above equation (56) with a respect to $\lambda_a$ we have the $T_{I_4}$ .

Using Bethe Integral [3] we get, equation (51) can be written as

$$T_{I_5} = -\frac{8\pi\lambda_a ie^{\frac{\pi\alpha_b}{2}}\Gamma(1-i\alpha_b)4\pi}{81\sqrt{6}\pi^2\{\lambda_a{}^2+(\bar{t}-\bar{p}_a)^2\}^2(t^2)^{1+i\alpha}\left(p_i^2+\bar{p}_i.\bar{p}_a+\bar{p}_i.\bar{p}_b+\bar{p}_a.\bar{p}_b\right)^{-i\alpha}} \tag{57}$$

Differentiating the above equation (57) with a respect to $\lambda_a$ we have the $T_{I_5}$

Again, using Bethe Integral [3] equation (52) can be written as

$$T_{I_6} = -\frac{ie^{\frac{\pi\alpha_b}{2}}(e^{\pi\alpha}-e^{-\pi\alpha})}{81\sqrt{6}\pi^3\Gamma(i\alpha)\Gamma(1-i\alpha)}\left[\int_0^1 dx\, x^{i\alpha}(1-x)^{-i\alpha}\Gamma(1-i\alpha)\left\{\frac{\phi_a(x)-\phi_a(o)}{x}\right\} - i\Gamma(1-i\alpha)\phi_a(o)\right] \tag{58}$$

Differentiating the above equation (58) with a respect to $\lambda_a$ we have the $T_{I_6}$.

$$\bar{q}'_b = -\bar{p}_a - \frac{x\bar{p}}{2}, \bar{q}'_a = -\left(\bar{p}_i - \bar{p}_b + \frac{x\bar{p}}{2}\right), \lambda_a = \frac{1}{3}, \lambda_c = \frac{xp}{2}$$

Putting all values of $T_{I_1}$, $T_{I_2}$, $T_{I_3}$, $T_{I_4}$, $T_{I_5}$ and $T_{I_6}$ in equation (46) we get the value of $T_I$.

The last term of equation (16) can be written as follows:

$$T_{PB} = \frac{1}{162\sqrt{6}\pi^2} \langle e^{-i\bar{p}_a \cdot \bar{\gamma}_a} e^{-i\bar{p}_b \cdot \bar{\gamma}_b} \left| \frac{1}{\gamma_{ab}} - \frac{Z}{\gamma_b} \right| e^{i.\bar{p}_i \cdot \bar{\gamma}_b} (27 - 18\gamma_a + 2\gamma_a^2) e^{-\lambda_a \cdot \gamma_a} \rangle$$

$$= \frac{1}{162\sqrt{6}\pi^2} \int e^{-i\bar{p}_a \cdot \bar{\gamma}_a} e^{-i\bar{p}_b \cdot \bar{\gamma}_b} \left| \frac{1}{\gamma_{ab}} - \frac{1}{\gamma_b} \right| (27 - 18\gamma_a + 2\gamma_a^2) e^{i\bar{p}_i \cdot \bar{\gamma}_b} \, e^{-\lambda_a \cdot \gamma_a} d^3\gamma_a d^3\gamma_b$$

$$T_{PB} = T_{PB_1} + T_{PB_2} + T_{PB_3} + T_{PB_4} + T_{PB_5} + T_{PB_6} \quad (59)$$

where, $T_{PB_1} = \frac{1}{6\sqrt{6}\pi^2} \int e^{-i\bar{p}_a \cdot \bar{\gamma}_a} e^{-i\bar{p}_b \cdot \bar{\gamma}_b} \frac{1}{\gamma_{ab}} e^{i\bar{p}_i \cdot \bar{\gamma}_b} e^{-\lambda_a \cdot \gamma_a} d^3\gamma_a d^3\gamma_b$ (60)

$$T_{PB_2} = -\frac{1}{9\sqrt{6}\pi^2} \int e^{-i\bar{p}_a \cdot \bar{\gamma}_a} e^{-i\bar{p}_b \cdot \bar{\gamma}_b} \frac{\gamma_a}{\gamma_{ab}} e^{i\bar{p}_i \cdot \bar{\gamma}_b} e^{-\lambda_a \cdot \gamma_a} d^3\gamma_a d^3\gamma_b \quad (61)$$

$$T_{PB_3} = \frac{1}{81\sqrt{6}\pi^2} \int e^{-i\bar{p}_a \cdot \bar{\gamma}_a} e^{-i\bar{p}_b \cdot \bar{\gamma}_b} \frac{{\gamma_a}^2}{\gamma_{ab}} e^{i\bar{p}_i \cdot \bar{\gamma}_b} e^{-\lambda_a \cdot \gamma_a} d^3\gamma_a d^3\gamma_b \quad (62)$$

$$T_{PB_4} = -\frac{1}{6\sqrt{6}\pi^2} \int e^{-i\bar{p}_a \cdot \bar{\gamma}_a} e^{-i\bar{p}_b \cdot \bar{\gamma}_b} \frac{1}{\gamma_b} e^{i\bar{p}_i \cdot \bar{\gamma}_b} e^{-\lambda_a \cdot \gamma_a} d^3\gamma_a d^3\gamma_b \quad (63)$$

$$T_{PB_5} = \frac{1}{9\sqrt{6}\pi^2} \int e^{-i\bar{p}_a \cdot \bar{\gamma}_a} e^{-i\bar{p}_b \cdot \bar{\gamma}_b} \frac{\gamma_a}{\gamma_b} e^{i\bar{p}_i \cdot \bar{\gamma}_b} e^{-\lambda_a \cdot \gamma_a} d^3\gamma_a d^3\gamma_b \quad (64)$$

$$T_{PB_6} = -\frac{1}{81\sqrt{6}\pi^2} \int e^{-i\bar{p}_a \cdot \bar{\gamma}_a} e^{-i\bar{p}_b \cdot \bar{\gamma}_b} \frac{{\gamma_a}^2}{\gamma_b} e^{i\bar{p}_i \cdot \bar{\gamma}_b} e^{-\lambda_a \cdot \gamma_a} d^3\gamma_a d^3\gamma_b \quad (65)$$

Using Bethe Integral [3] equation (60) becomes

$$T_{PB_1} = -\frac{2*\lambda_a}{3\sqrt{6}t^2\{\lambda_a^2 + (\bar{t} - \bar{p}_a)^2\}^2} \quad (66)$$

Differentiating the above equation (66) with a respect to $\lambda_a$, we get term $T_{PB_1}$. Similarly, Using Bethe Integral [3] equation (61) becomes,

$$T_{PB_2} = \frac{16*\lambda_a}{3\sqrt{6}t^2\{\lambda_a^2+p_a^2\}^2} \qquad (67)$$

Differentiating the above equation (67) with a respect to $\lambda_a$, we get term $T_{PB_2}$. Bethe Integral [3] the equation (62) gives

$$T_{PB_3} = \frac{32*[3\lambda_a^2-(\bar{t}-\bar{p}_a)^2]}{9\sqrt{6}t^2\{\lambda_a^2+(\bar{t}-\bar{p}_a)^2\}^2} \qquad (68)$$

Differentiating the above equation (68) with a respect to $\lambda_a$ we have the $T_{PB_3}$, Using Bethe Integral [3] equation (63) becomes

$$T_{PB_4} = -\frac{32*\lambda_a}{9\sqrt{6}t^2(\lambda_a^2+p_a^2)^2} \qquad (69)$$

Differentiating the above equation (69) with a respect to$\lambda_a$ we have the $T_{PB_4}$.

Using Bethe Integral [3] equation (64) becomes

$$T_{PB_5} = \frac{32*[3\lambda_a^2-(\bar{t}-\bar{p}_1)^2]}{9\sqrt{6}t^2\{\lambda_a^2+(\bar{t}-\bar{p}_1)^2\}^2} \qquad (70)$$

Differentiating the above equation (70) with a respect to $\lambda_a$ we have the $T_{PB_5}$

Using Bethe Integral [3] equation (65) becomes

$$T_{PB_6} = \frac{32*(3\lambda_a^2-p_a^2)}{9\sqrt{6}t^2(\lambda_a^2+p_a^2)^2} \qquad (71)$$

After differentiating the above equation (71) with a respect to $\lambda_a$ we have the $T_{PB_6}$.

Putting all the values of in equation (59) we get the final term of $T_{PB}$. The direct scattering amplitude $F(\bar{p}_a, \bar{p}_b)$ is then determined from

$$F(\bar{p}_a, \bar{p}_b) = -(2\pi)^2 T_{FI} \tag{72}$$

We have also calculated the above equations analytically for our current study using the Lewis Integral [20]. Now, the Triple Differential Cross Section (TDCS) corresponding to the T-matrix element is given by:

$$\frac{d^3\sigma}{dE_a d\Omega_a\, d\Omega_b} = \frac{p_a\, p_b}{p_i} |T_{FI}|^2 \tag{73}$$

Now the double differential cross section (DDCS) is obtained by integrating [8] equation (73) with respect to solid angle $\Omega_b$.

$$\frac{d^2\sigma}{dE_a\, d\Omega_a} = \int \frac{d^3\sigma}{dE_a\, d\Omega_a\, d\Omega_b} d\Omega_b \tag{74}$$

Hence, the resulting expressions were numerically computed using a programming language MATLAB.

## 3. Results and Discussions

The double differential cross sections (DDCS) for the ionization of meta stable hydrogen in the 3S state due to electron impact have been computed for ejected electron energies $E_a$ at incident energies $E_I$ = 250eV and $E_I$= 150eV. The results have been graphed as a function of the ejection angle $\theta_a$(0°–360°) for constant scattered electron geometries (scattering angle $\theta_b$, 0° up to 90°). In the main figures, $\theta_a$ is depicted within the range of 0°–60° to emphasize the predominant recoil and binary configurations. The current DDCS have been juxtaposed with experimental ground state data and the first-Born results from Shyn [2], as well as the theoretical calculations by Das et al. [1]; the first-Born curves have been graphically shown for direct comparison. The recoil zone is defined by $\theta_a$= 0°–90° at $\phi$ = 0°, while the binary region is characterized by $\theta_a$= 90°–180° at $\phi$ = 180°.

**Figure 1.** Illustrates the Double Differential Cross Section (DDCS) for an incident energy ($E_I$) of 250eV and an energy transfer ($E_a$) of 4eV. The current second-born outcome exhibits a smooth forward (recoil) peak and a significant decrease at greater θ. Qualitatively, this scenario resembles the experimental framework of Shyn [2], and the current DDCS aligns more closely with the ground-state measurements of Shyn [2] than with the calculations of Das et al. [1] in the recoil area. The firstborn exhibits a comparable overall pattern, accompanied by slight quantitative variations. The positron impact differential double cross section (DDCS) exhibits two distinct maxima corresponding to the recoil and binary lobes, with positron results surpassing those of the electron DDCS at about $\theta_a$= 0°–100° and $\theta_a$= 90°–160°.

**Figure 2.** Illustrates the Double Differential Cross Section (DDCS) for an incident energy ($E_I$) of 250eV and an energy transfer ($E_a$) of 10eV. The current 3S computation and the findings of Das et al. (1995) exhibit comparable patterns and are likely in satisfactory concordance, especially in the binary region. At elevated ejection angles, Shyn's results align more closely with the current calculations than with those of Das et al. Furthermore, the current DDCS is marginally greater than that of Das and Shyn in the binary region; in the recoil region, the present computation complements the two references. The current second-born and first-born DDCS are nearly indistinguishable through this kinematics.

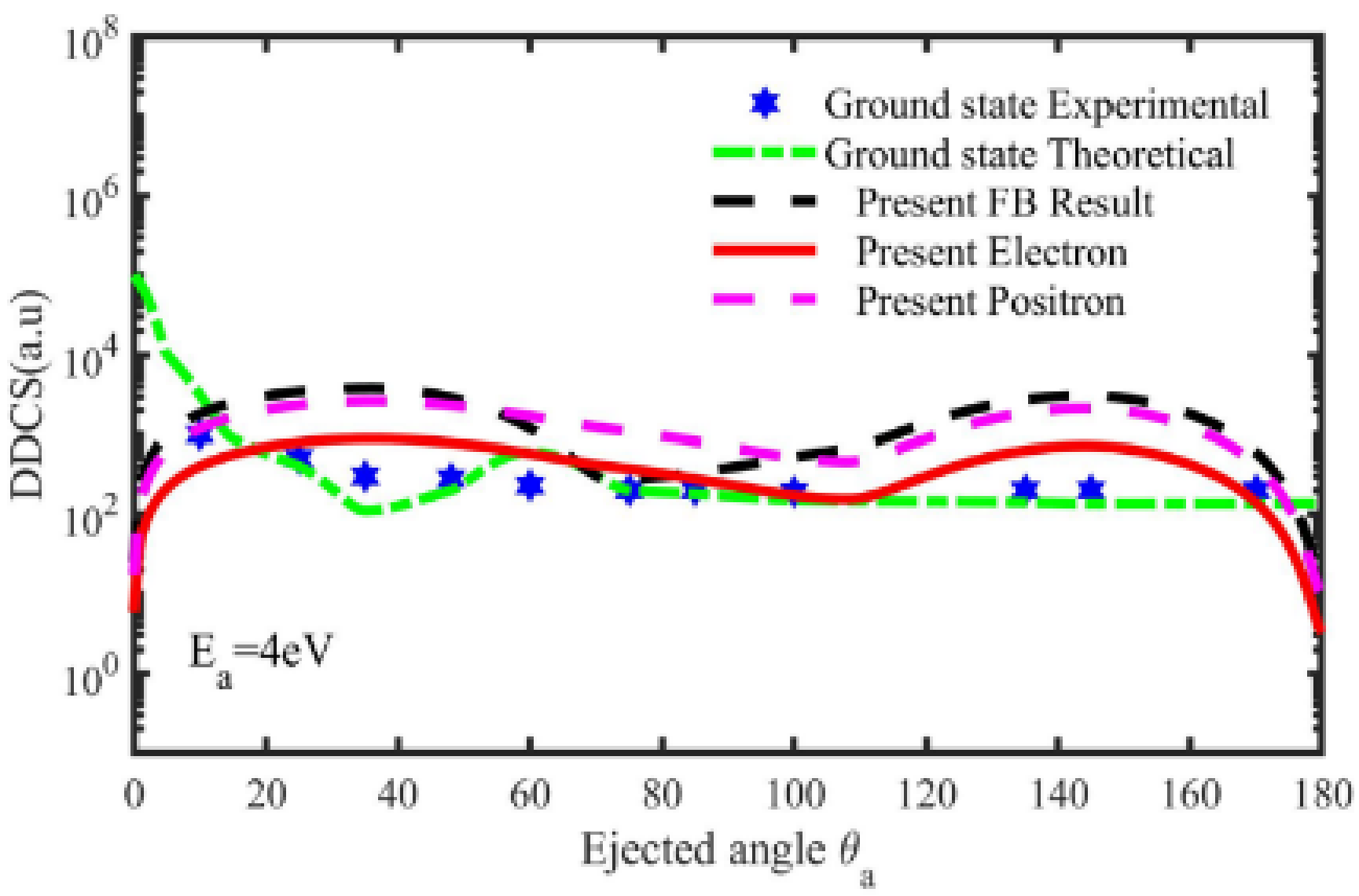

**Figure 1:** Second Born double differential cross section (DDCS) as a function of electron impact energy $E_I = 250\text{eV}$ and ejected electron energy $E_a = 4\text{eV}$.

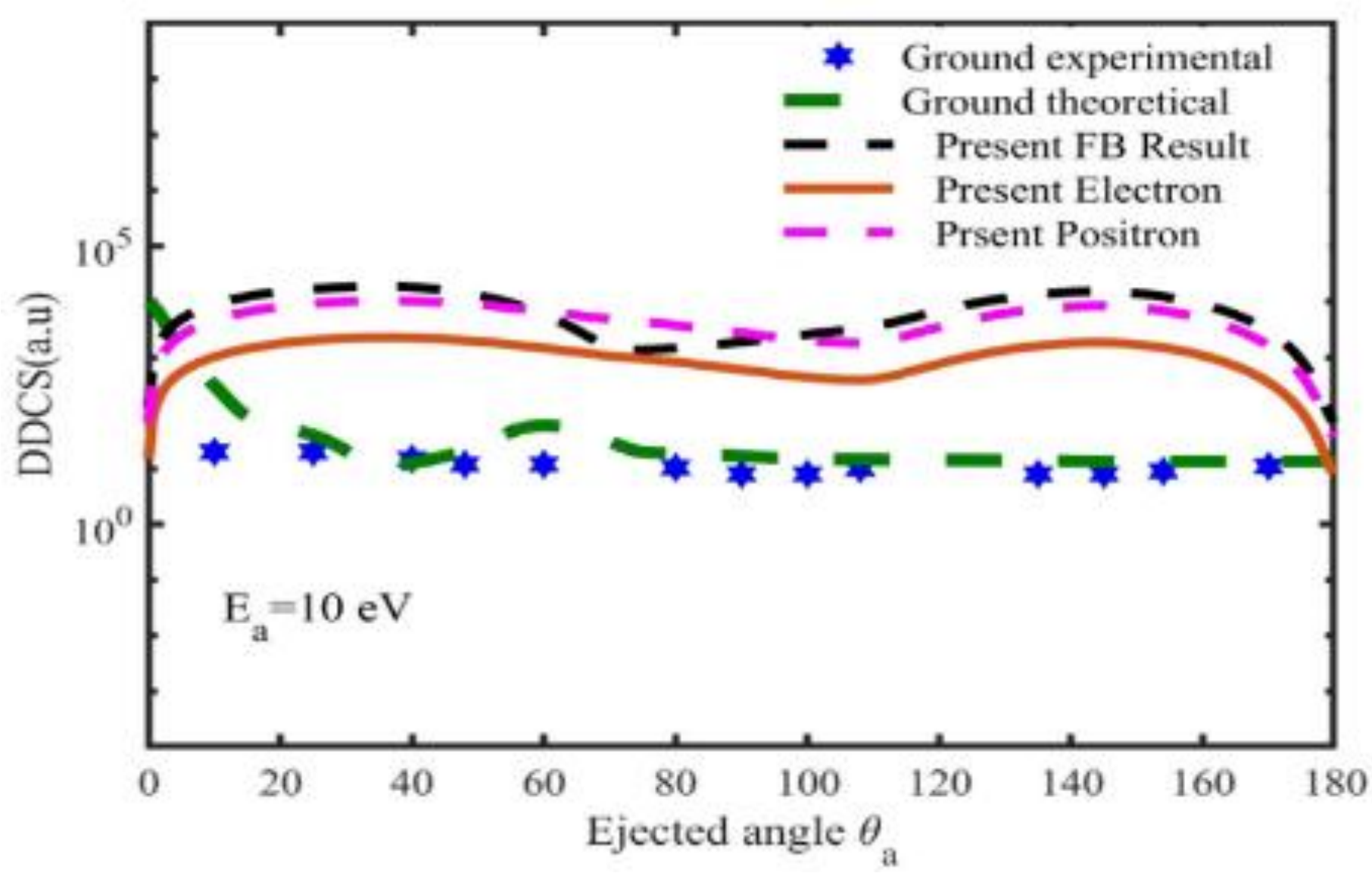

**Figure 2:** Second Born double differential cross section (DDCS) for electron impact energy $E_I = 250eV$ and ejected electron energy $E_a = 10eV$.

Additionally, we illustrate the initial results of the incident electron energy and the ejected electron energy in Figure 3; the two peaks corresponding to the recoil and binary regime structures are evident in the initial results, and the findings for high ejected energy are in strong agreement with the experimental results. The residual computations and the hydrogen state outcomes from Das et al. [1] are congruent. Conversely, other experimental data exhibit an opposing profile in the recoil region; nonetheless, the current results align effectively with Shyn's experimental findings [2] in the binary region, suggesting that the present computations hold greater significance under the kinematic conditions.

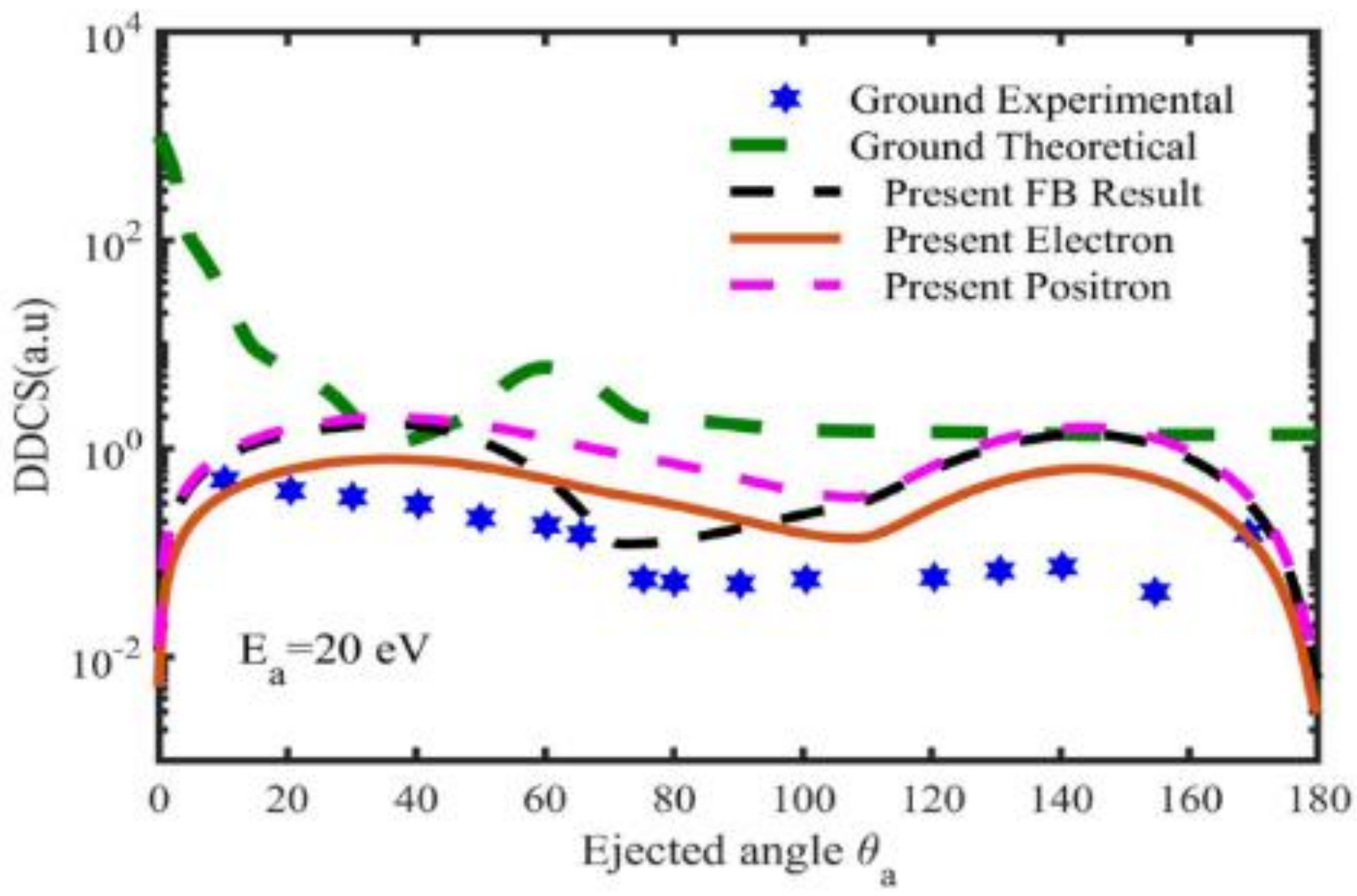

**Figure 3:** Second Born double differential cross section (DDCS) for the electron impact energy $E_I = 250eV$ and the ejected electron energy $E_a = 20eV$.

The findings indicate an excitation energy of 250eV, aligning well with all differential cross-section measurements and studies. Furthermore, all computed DDCS results for the metastable 3S-state closely align with Shyn's [2] experimental findings in the binary zone, particularly evident in this kinematic context. The hydrogen second Born result exhibits notable discrepancies when juxtaposed with the findings for the hydrogen ground state [1], [2]. This outcome pertains to the kinematics of two-body electron-electron interactions and the final state interaction between the electron and nucleus, resulting in a broad peak-shaped pattern in the double differential cross sections for the ejected electron at an incident energy of 250eV following electron impact. In summary, this strategy has yielded results that are qualitatively comparable to prior findings on a hydrogen ground state.

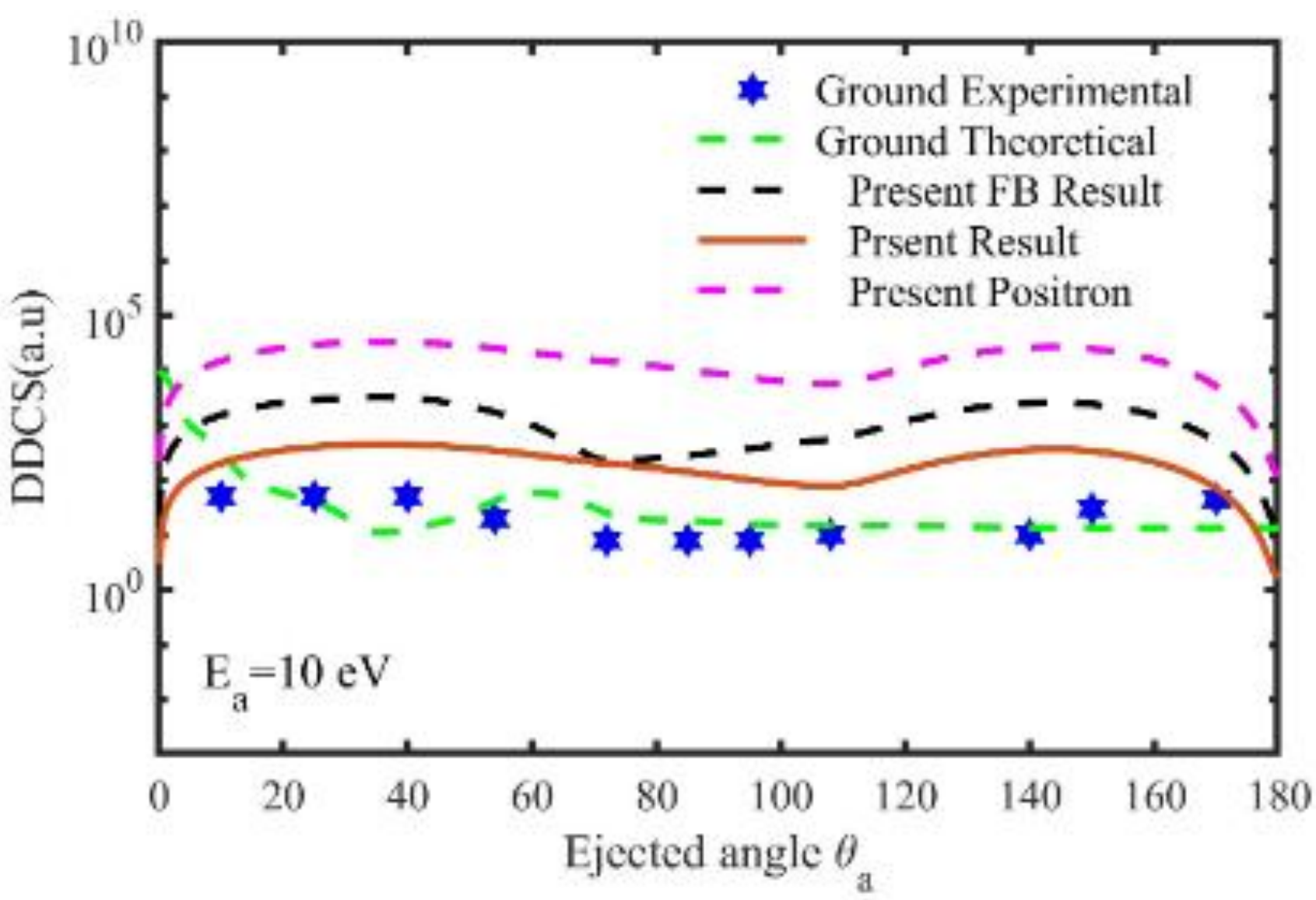

**Figure 4:** The Second Born double differential cross section (DDCS) for electron impact energy $E_I = 150eV$ and ejected electron energy $E_a = 10eV$.

Final Assessment This section aims to examine the incident energy value $E_I$=150eV for various ejected energies $E_a$= 10eV, 20eV, and 50eV, as illustrated in Figure 4, Figure 5, and Figure 6.

Initially, in Figure 5, our findings for incident energy, $E_I$=150eV and $E_a$=20eV, align qualitatively with the hydrogen ground state [3], and [4], given that the ejection angle exceeds 50°. Subsequently, it ascends significantly higher than both comparisons and has a smooth apex in binary rationale. The data for first-born events also exhibit the same overall trend, with the sole distinction being the presence of dips in the recoil zone. In positron impact, particularly at lower energies, the combined results of current and first-born calculations align closely with the hydrogen ground state findings at both small and large angles.

Additionally, in the present recoil zone, there appears to be no smooth bump, although it is not evident in the current findings but rather in more efficient ones. Furthermore, in the binary peak region, it appears to decelerate more with for an ejected electron energy of $E_I$=10eV and an impact energy of $E_I$=150eV in Figure 6, pertaining to the recoil area, and in conjunction with Shyn [4], at an elevated ejection angle, alongside the hydrogen ground state data [3], and [4]. The computations demonstrate remarkable concordance for the initial measurement, a robust correlation across all comparative measurements, and the recoil lobe is accurately depicted in the angular distributions. The second-born and first-born DDCS align closely with the reference curves, and the final results exhibit consistent behaviour relative to the examined kinematic conditions, ensuring qualitative fidelity between the theoretical predictions and the available experimental data for comparison.

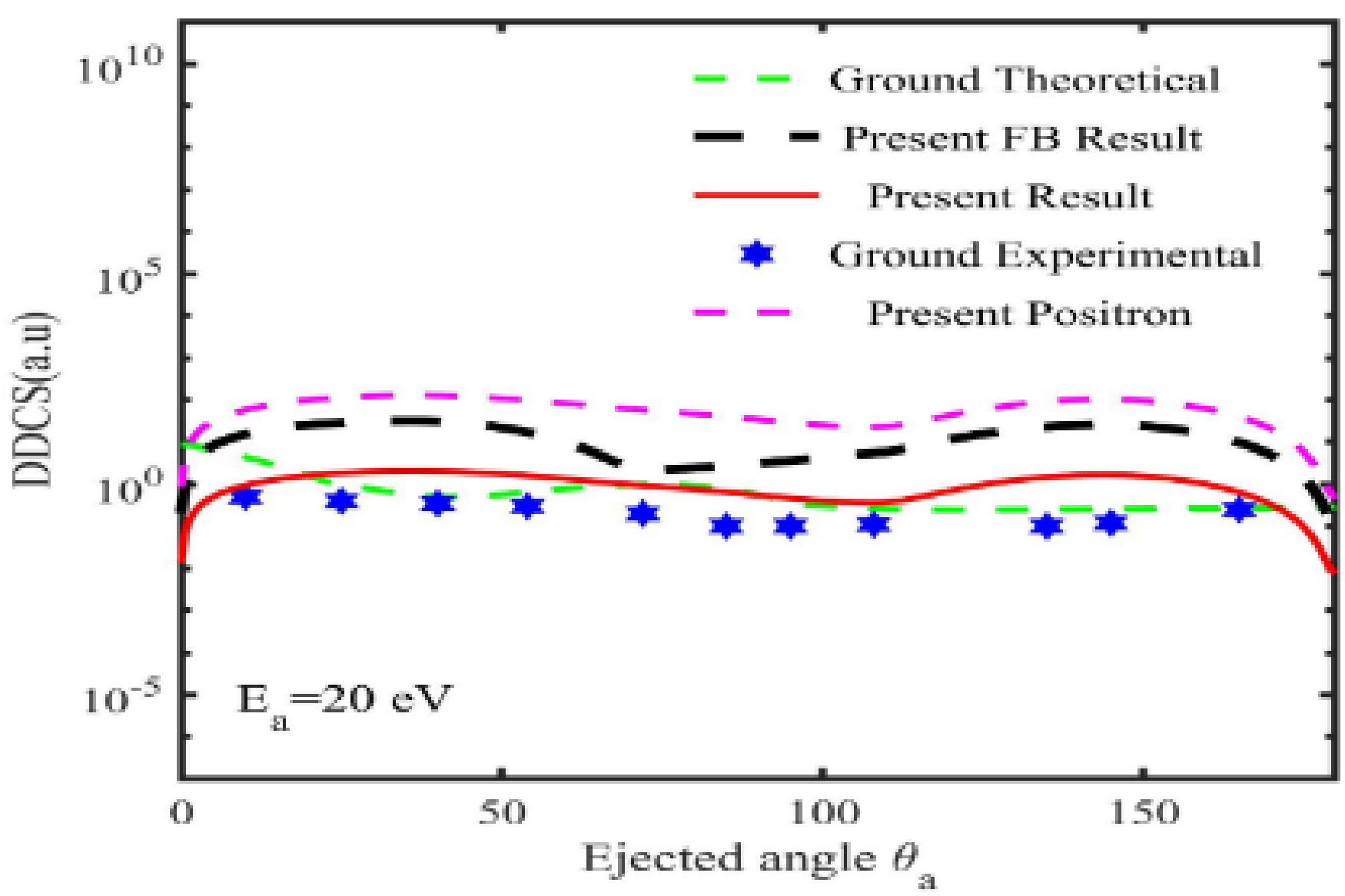

**Figure 5:** Second Born double differential cross section (DDCS) for electron impact energy $E_I = 150eV$ and ejected electron energy $E_1 = 20eV$.

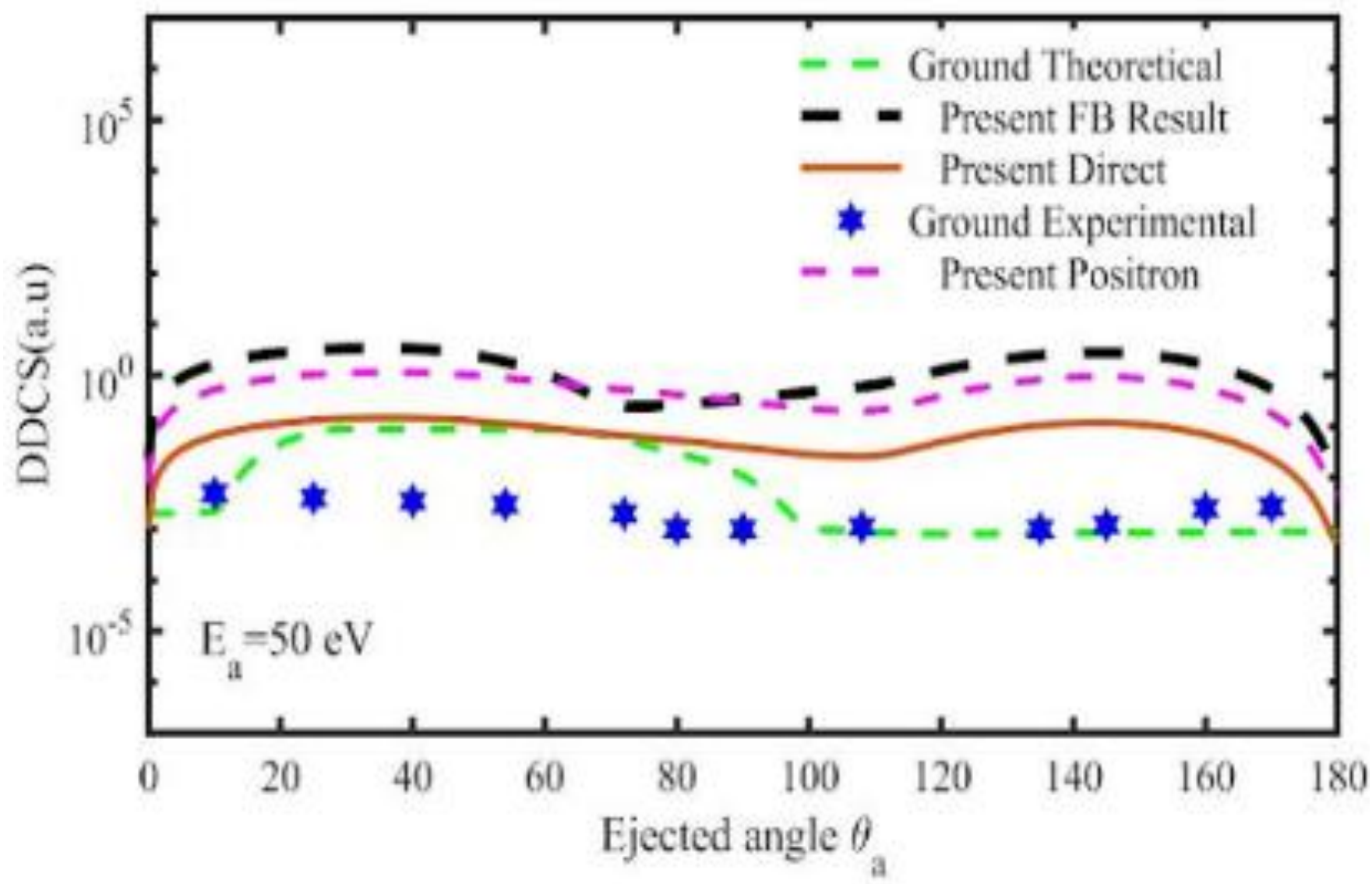

**Figure 6:** Second Born double differential cross section (DDCS) for electron impact energy $E_I = 150eV$ and ejected electron energy $E_a = 50eV$.

To understand these structures of the DDCS results one may look carefully to the Table-1.

| $\theta_a$ (deg) | $\theta_b$(deg) | DDCS $E_e$=4eV | DDCS $E_e$=10eV | DDCS $E_e$=20eV |
|---|---|---|---|---|
| 0 | 0 | 0.0058E-03 | 0.0149E-04 | 0.0052 |
| 1 | 36 | 0.8886E-03 | 2.2734E-04 | 0.7917 |
| 2 | 72 | 0.3972E-03 | 1.0161E-04 | 0.3539 |
| 4 | 108 | 0.1539E-03 | 0.3937E-04 | 0.1371 |
| 10 | 144 | 0.7179E-03 | 1.8366E-04 | 0.6396 |
| 20 | 180 | 0.0032E-03 | 0.0083E-04 | 0.0029 |
| 30 | 216 | 1.0109E-03 | 2.5864E-04 | 0.9007 |
| 40 | 252 | 0.0161E-03 | 0.0412E-04 | 0.0143 |
| 60 | 288 | 0.2904E-03 | 0.7431E-04 | 0.2588 |
| 90 | 324 | 0.5271E-03 | 1.3485E-04 | 0.4696 |
| 100 | 360 | 0.2077E-03 | 0.5313E-04 | 0.1850 |

**Table 1:** DDCS results for ejected angles $\theta_a$ corresponding to various scattering angles $\theta_b$ for $E_a$=4eV , $E_a$=10eV , $E_a$=20eV in ionization of hydrogen atoms for $E_I = 250eV$.

Furthermore, they augment the energy expelled, ultimately resulting in $E_a$=50eV, while the incident energy $E_I$=150eV remains constant. The DDCS outcome from a metastable 3S state is further advanced by Das et al. [3] in the recoil area. It is also one region earlier than Shyn [2] in the recoil region, exhibiting a tendency that is largely contrary to that of Shyn [2] for the same area, and it merges with Shyn [2] in the binary region. The first-born estimate closely aligns with [1] and [2] for lower-angle emission, as well as with the current data and Shyn's findings [2] for higher-angle emission. At the recoil limit, our results exhibit favourable qualitative alignment with the ground state hydrogen outcomes presented in [1] and [2].

However, there exists a significant disparity with the larger zero values, despite the first-born approximation demonstrating reasonable concordance, approximately within five significant figures. However, we do note the distinctions arising from the quantity of hydrogen atomic states, and our current research may assert a strong qualitative equivalence to both hydrogen ground states, as demonstrated.

## 4. Conclusion

We have investigated the double differential cross sections (DDCS) for the ionization of a metastable hydrogen atom (3S state) after electron impact using the theory of Das et al. [1]. The current results have very broad and smooth forward peaks for larger ejection angles in a similar fashion to the trends noted by Shyn [2], Bethe [3], and Das et al. [1], while the results from calculations with hydrogen ground state aspects showed greater differences. We find reasonably much greater differences for the small ejection angles. However, for most cases the second-born DDCS calculations presented in the current study are still qualitatively comparable to the previous studies that investigated the hydrogen atom in the ground state. In total, these results provide an important theoretical advance in our understanding of the ionization of a hydrogen atom in the excited state, i.e., the metastable 3S state.

Theoretical DDCS cannot be compared directly against experimental results, as there are currently no DDCS results for the meta-stable 3S state; therefore, the current calculation provides theoretical benchmarks for future experimental work. The current work could also be expanded to include a broader range of incident/ejected energies, other hydrogen meta-stable states, and other ionization processes that can result from positron impact. The current predictions could be improved further with the inclusion of relativistic effects and higher-order coupling effects, which would allow us to develop a deeper understanding of electron and atom collision dynamics in an excited system.

## 5. Acknowledgements

The computations were done in the Simulation Lab, Department of Mathematics, Chittagong University of Engineering and Technology, Chittagong, 4349, Bangladesh.

## 6. Conflicts of interest

The authors declare that they have no conflicts of interest regarding the publication of this paper.